\documentstyle[12pt,epsfig,axodraw]{article}

\newcommand{\lsim}{\raisebox{-0.13cm}{~\shortstack{$<$ \\[-0.07cm] $\sim$}}~}
\newcommand{\gsim}{\raisebox{-0.13cm}{~\shortstack{$>$ \\[-0.07cm] $\sim$}}~}

\newcommand{\ee}{e^+e^-}
\newcommand{\s}{   }
\newcommand{\nn}{\noindent}
\newcommand{\non}{\nonumber}
\newcommand{\beq}{\begin{eqnarray}}
\newcommand{\eeq}{\end{eqnarray}}
\newcommand{\tb}{\tan\beta}

\begin{document}

\def\thefootnote{\fnsymbol{footnote}}

\begin{flushright}
PM/01--06\\
January 2001\\
\end{flushright}

\vspace{1cm}

\begin{center}

{\large\sc {\bf Probing the SUSY Higgs boson couplings to scalar leptons}}

\vspace{.4cm}

{\large\sc {\bf at high--energy e$^+$e$^-$ colliders}}

\vspace{1cm}

{\sc Aseshkrishna Datta}, {\sc Abdelhak Djouadi} and {\sc Jean-Lo\"{\i}c Kneur} 

\vspace{0.5cm}

Laboratoire de Physique Math\'ematique et Th\'eorique, UMR5825--CNRS,\\
Universit\'e de Montpellier II, F--34095 Montpellier Cedex 5, France.

\end{center}

\vspace{2cm}

\begin{abstract}
\nn We discuss the production at $e^+e^-$ colliders of Higgs bosons in
association with both the scalar leptons and the lightest neutralinos in the
Minimal Supersymmetric extension of the Standard Model. While the rates for
associated Higgs production with neutralinos and first/second generation
sleptons are rather tiny, the cross section for the production of the lightest
Higgs boson $h$ with scalar $\tau$ lepton pairs can reach the femtobarn level
at c.m. energies at and above 500 GeV in favorable regions of the parameter 
space, making this process potentially detectable at a high--luminosity $\ee$
collider, in particular in the $\gamma \gamma$ option. This would provide a 
determination of the $h \tilde{\tau} \tilde{\tau}$ coupling and opens up the 
possibility of measuring the parameter $\tb$.
\end{abstract} 
 
\def\thefootnote{\arabic{footnote}}
\setcounter{footnote}{0}

\newpage

\subsection*{1. Introduction}

The search for Supersymmetry (SUSY) is one of the major goals of present and
future high--energy colliders. Once SUSY particles are found, it would be of
prime importance to study in detail their properties and interactions in order
to reconstruct the SUSY Lagrangian. This will be mandatory to decide which SUSY
scenario is effectively realized at the low energies probed by experiments and
potentially, to derive the structure of the theory at high scales.  

The SUSY Lagrangian can be reconstructed by measuring the couplings between the
SUSY and the standard particles. Among these, the couplings of sparticles to
Higgs bosons are of special importance since they also probe the electroweak
symmetry breaking sector and might decide which Higgs scenario is at work.
In the Minimal Supersymmetric Standard Model (MSSM) \cite{MSSM}, two Higgs
doublet fields \cite{HHG} are needed to break the SU(2)$\times$U(1) symmetry,
leading to a quintet of Higgs bosons, the lightest of which, the neutral
CP--even scalar $h$, should have a mass below $\sim 130$ GeV \cite{hmass}. The
couplings of the Higgs bosons to the SUSY scalar fermions $\tilde{f}$ and to
the charginos $\chi^\pm$ and neutralinos $\chi^0$ depend on the soft--SUSY
breaking parameters and therefore carry informations on the fundamental SUSY
theory.  

In the MSSM, the Higgs boson couplings to the charginos $\chi^\pm_{1,2}$ and
neutralinos $\chi^{0}_{1,\cdots,4}$ depend on $\tb$, the ratio of the vev's of
the two Higgs fields, the higgsino parameter $\mu$ and the bino and wino mass
parameters $M_1$ and $M_2$ which are linked by the relation $M_1=
\frac{5}{8}\tan^2 \theta_W M_2 \simeq \frac{1}{2} M_2$ in the minimal
Supergravity (mSUGRA) \cite{mSUGRA} model where the gaugino masses [as well as
the scalar masses and the trilinear sfermion couplings] are unified at the GUT
scale and where the electroweak symmetry is broken radiatively. For instance,
the $h \chi_1^0 \chi_1^0$ coupling which might be the first to be accessible
[since the $h$ boson is light and the neutralino $\chi_1^0$ is expected to be
the lightest sparticle (LSP) in the MSSM] is given, in the decoupling limit 
[where the $h$ boson becomes Standard Model like and all the other Higgs 
particles are heavy], by \cite{gunion}
\beq
g_{h \chi_i^0 \chi_j^0} \propto (Z_{i2} -\tan \theta_W Z_{i1})(\sin \beta
Z_{j3}+\cos \beta Z_{j4}) + i \leftrightarrow j 
\eeq
with ${\small i=j=1}$. Here $Z_{ij}$ are the elements of the matrix $Z$ which
diagonalises the $4 \times 4$ neutralino mass matrix. As
can be seen, the $h$ boson couples to mixtures of gaugino $(Z_{i1}, Z_{i2})$
and higgsino $(Z_{i3}, Z_{i4})$ components of the neutralinos. If the light
neutralino $\chi_1^0$ were a pure bino [as is the case in a large part of the 
mSUGRA model parameter space \cite{DM}, in particular when cosmological 
constraints are incorporated] or a pure higgsino, the coupling would vanish 
and thus would be hard to measure experimentally. 

The couplings of the $h$ boson to sfermion pairs, $g_{h\tilde{f}_i\tilde{f}_j}$,
can be stronger. In the decoupling limit, and in terms of $\tb$, $\mu$
and the trilinear coupling $A_f$, the (normalized) diagonal and non--diagonal
$h$--sfermion couplings read in the MSSM [$s_{\theta_f}=\sin \theta_f, 
c_{\theta_f}=\cos \theta_f$, etc ...] 
\beq
\bigg( \begin{array}{c} g_{h \tilde{f}_1 \tilde{f}_1 } \\ g_{h \tilde{f}_2 
\tilde{f}_2} \end{array} \bigg) &=& \cos 2\beta \left[I_f^3 
\left( \begin{array}{c} c^2_{\theta_f} \\ s^2_{\theta_f} \end{array} \right)
- e_f s^2_W c_{2 \theta_f} \right] + \frac{m_f^2}{M_Z^2} \pm \frac{s_{2
\theta_f} m_f}{2M_Z^2} [A_f -\mu (\tb)^{-2I_f^3} ] \non \\
g_{h \tilde{f}_1 \tilde{f}_2 } &=& \cos 2\beta \;  s_{2\theta_f} [e_f s_W^2 
-  I_f^3/2] + c_{2\theta_f} m_f [ A_f -\mu (\tb)^{-2I_f^3}]/(2M_Z^2) 
\eeq
where $I_f^3$ is the weak isospin and $e_f$ the electric charge of the sfermion
$\tilde{f}$ and $\theta_f$ the mixing angle between the left and right--handed
sfermions $\tilde{f}_L$ and $\tilde{f}_R$ [which as for the sfermion masses 
$m_{\tilde{f}_1}$ and $m_{\tilde{f}_2}$, are given in terms of the three
parameters above and the soft--SUSY breaking scalar masses $m_{\tilde{f}_L}$
and $m_{\tilde{f}_R}$]; $s_W^2=1-c_W^2 \equiv \sin^2 \theta_W$.  For first and
second generation sfermions, as apparent from eq.~(2), these couplings are
relatively tiny since $m_f$ and the mixing angle $\theta_f$ are small, and the
term proportional to $\cos2\beta$ is not enhanced.  

The couplings of Higgs bosons to the third generation squarks, $\tilde{t}$ and
$\tilde{b}$, can be much larger and potentially measurable in the associated
Higgs+squark production process at proton or $e^+e^-$ colliders, as discussed
in Refs.~\cite{hstop0, hstop}.  In this paper, we will investigate the
prospects of measuring the lightest CP--even Higgs boson couplings to sleptons
which can be best performed in the clean environment of future high--energy
$e^+e^-$ colliders\footnote{In this paper, we will not discuss the case of the
heavier MSSM Higgs bosons since the associated production processes are less
favored by phase space. In addition, we will stick to $\ee$ colliders, since
at hadron colliders, the cross section for (electroweak) Higgs--slepton 
production is relatively much smaller than the potential
backgrounds, and the signal would be more complicated to extract.} \cite{eeH}.
Contrary to the squark case, the measurement of these couplings can be
performed in two ways: 

$i)$ In the production of LSP pairs, $\ee \to \chi_1^0 \chi_1^0$, which is the 
first kinematically accessible SUSY process in $\ee$ collisions, the $h$ boson 
can be emitted not only for the final $\chi_1^0$ lines but also from the 
selectrons which are exchanged in the $t$ and $u$ channels. The production 
cross section for the $\ee \to \chi_1^0 \chi_1^0 h$ associated process thus
involves the $h \tilde{e} \tilde{e}$ couplings. 

$ii)$ In the production of selectrons $\ee \to \tilde{e} \tilde{e}^*$ or
sneutrinos  $\ee \to \tilde{\nu}_e \tilde{\nu}_e^*$, the Higgs bosons can be
emitted from both the final slepton lines or from, respectively, the
neutralinos and charginos which are exchanged in the $t$--channels; the
production cross sections are then in principle proportional to complicated
combinations of the Higgs couplings to sleptons and neutralino/chargino states.
In the case of smuons and staus and their corresponding sneutrinos,  there is
no gaugino exchange channels and the processes $\ee \to \tilde{l}\tilde{l}^*h$
are mediated by $s$--channel $(\gamma)Z$ exchange with the Higgs boson emitted
from the slepton lines.  Up to the small contribution of the diagrams where the
$h$ boson is emitted from the $Z$--boson line [see later] the cross sections are
directly proportional to the square of the $h \tilde{l} \tilde{l}$ couplings
which would be then, in principle, measurable in these processes. \s

In this paper we analyze the prospects of measuring the Higgs--slepton
couplings at high--energy and high--luminosity $\ee$ colliders.  In Sections 2
and 3, we discuss the associated production of the $h$ boson with,
respectively, the lightest neutralinos and selectron/sneutrino states.  In
Section 4 we focus on the case of stau leptons where the cross sections, in
both $e^+e^-$ and $\gamma \gamma$ options of the $\ee$ collider,
will be shown to be potentially large.  Some conclusions will be given in the
final Section 5.  
 
\subsection*{2. Higgs boson production in association with neutralinos}

The Feynman diagrams contributing to the production of the lightest CP--even
Higgs boson in association with neutralino pairs is shown in Fig.~1 [the
diagrams where the $h$ boson is emitted from the electron and positron lines
give negligible contributions]. A first class of contributions (1a) is formed
by diagrams where the Higgs boson is emitted from the neutralino states, the
latter being produced through $s$--channel $Z$ boson exchange and $t$--channel
left-- and right--handed selectron exchange. A second class (1b) is formed by
the Higgs--strahlung production process, where the $Z$ boson is virtual and
splits into two neutralinos. Finally, a third class (1c) consists of the
diagrams where the Higgs boson is emitted from the internal selectron lines.
The cross section will therefore depend on the $h$ boson couplings to both the
neutralinos and sleptons.  

\vspace*{.7cm}
\begin{picture}(1000,180)(0,0)
\Text(0,180)[]{${\bf a)}$}
\Text(15,130)[]{$e^+$}
\Text(15,170)[]{$e^-$}
\ArrowLine(20,120)(40,150)
\ArrowLine(20,180)(40,150)
\Photon(40,150)(80,150){4}{6}
\Text(60,163)[]{$Z$}
\ArrowLine(80,150)(100,120)
\ArrowLine(80,150)(100,180)
\Text(105,120)[]{$\chi$}
\Text(105,180)[]{$\chi$}
\DashLine(92,170)(110,150){4}
\Text(105,145)[]{$h$}
\Text(135,130)[]{$e^+$}
\Text(135,170)[]{$e^-$}
\ArrowLine(140,120)(160,150)
\ArrowLine(140,180)(160,150)
\Photon(160,150)(200,150){4}{6}
\Text(180,163)[]{$Z$}
\ArrowLine(200,150)(220,120)
\ArrowLine(200,150)(220,180)
\Text(225,120)[]{$\chi$}
\Text(225,180)[]{$\chi$}
\DashLine(213,130)(230,150){4}
\Text(225,155)[]{$h$}
\Text(255,130)[]{$e^+$}
\Text(255,170)[]{$e^-$}
\ArrowLine(250,120)(290,120)
\ArrowLine(250,180)(290,180)
\DashArrowLine(290,120)(290,180){4}
\Text(280,150)[]{$\tilde{e}$}
\ArrowLine(290,120)(330,120)
\ArrowLine(290,180)(330,180)
\Text(330,130)[]{$\chi$}
\Text(330,170)[]{$\chi$}
\DashLine(310,180)(330,150){4}
\Text(320,155)[]{$h$}
\Text(355,130)[]{$e^+$}
\Text(355,170)[]{$e^-$}
\ArrowLine(350,120)(390,120)
\ArrowLine(350,180)(390,180)
\DashArrowLine(390,120)(390,180){4}
\Text(380,150)[]{$\tilde{e}$}
\ArrowLine(390,120)(430,120)
\ArrowLine(390,180)(430,180)
\Text(430,130)[]{$\chi$}
\Text(430,170)[]{$\chi$}
\DashLine(410,120)(430,150){4}
\Text(415,145)[]{$h$}
\Text(25,80)[]{${\bf b)}$}
\Text(55,30)[]{$e^+$}
\Text(55,70)[]{$e^-$}
\ArrowLine(60,20)(90,50)
\ArrowLine(60,80)(90,50)
\Photon(90,50)(130,50){4}{6}
\Text(110,63)[]{$Z$}
\DashLine(130,50)(160,20){4}
\Text(165,30)[]{$h$}
\Photon(130,50)(150,70){4}{6}
\ArrowLine(150,70)(180,90)
\ArrowLine(150,70)(180,60)
\Text(185,90)[]{$\chi$}
\Text(185,60)[]{$\chi$}
\Text(250,80)[]{${\bf c)}$}
\Text(275,30)[]{$e^+$}
\Text(275,70)[]{$e^-$}
\ArrowLine(270,20)(330,20)
\ArrowLine(270,80)(330,80)
\DashLine(330,20)(330,80){4}
\Text(320,50)[]{$\tilde{e}$}
\ArrowLine(330,20)(390,20)
\ArrowLine(330,80)(390,80)
\Text(390,30)[]{$\chi$}
\Text(390,70)[]{$\chi$}
\DashLine(330,50)(390,50){4}
\Text(360,60)[]{$h$}
\Text(330,50)[]{$\bullet$}
\end{picture}

\nn Figure 1: Feynman diagrams contributing to the $\ee \to h\chi_1^0\chi_1^0$ 
production process.
\vspace*{4mm} 

If the higgsino mass parameter $\mu$ is much larger than the bino and wino mass
parameters $M_1$ and $M_2$, $|\mu| \gg M_1$, the lightest neutralino is a pure
bino and its mass is given by $m_{\chi_1^0} \simeq M_1$.  In this case, the
neutralino couplings to the $h$ boson, eq.~(1), as well as to the $Z$--boson,
$g^{L,R}_{\chi_i^0 \chi_j^0 Z} \propto (Z_{i3} Z_{j3} - Z_{i4} Z_{j4})$, 
are small. The only diagram which would contribute to the production
rate is then diagram (1c) where the Higgs boson is emitted from the selectron
lines. The cross section will then be proportional only to the couplings $g_{h
\tilde{e}\tilde{e}}$. For these couplings, one can set $m_e=0$ and vanishing
mixing angle and only the first term $\propto \cos 2\beta$ will be present. For
large values of $\tan \beta$ [which are required to maximize the $h$ boson mass
and to evade the experimental constraint from LEP2 searches \cite{lep2}, $M_h
\gsim 113.5$ GeV in the decoupling regime] $\cos 2\beta \to -1$ and one has
$g_{h \tilde{l}_i \tilde{l}_j} = - \delta_{ij} (I_l^3 - e_l s_W^2)/(s_Wc_W)$.
Note that because $s_W^2 \sim 1/4$, the $h$ couplings to left-- and
right--handed selectrons are almost equal [in absolute value] and equal to half
of the $h \tilde{\nu} \tilde{\nu}$ coupling.  [If the neutralino $\chi_1^0$
were a pure higgsino, i.e. $\mu \ll M_1$, the $h$--$\chi_1^0$--$\chi_1^0$ as
well as the $\tilde{e}$--$e$--$\chi_1^0$ couplings would vanish; the only
diagram which would contribute to the process $\ee \to h \chi_1^0 \chi_1^0$
would be the diagram (1b) which does not involve any Higgs coupling to
superparticles].  

We have calculated the cross section  $\sigma(\ee \to \chi_1^0 \chi_1^0 h)$ and
the results at $\sqrt{s}=500$ and 800 GeV, for selected values of the LSP and
selectron masses, are given in Table 1. We have assumed that the LSP neutralino
is a pure bino so that $m_{\chi_1^0}=M_1$ and $Z_{11}=1$ and we used the
approximation $m_{\tilde{e}_L}=m_{\tilde{e}_R} =m_{\tilde{e}}$; we have fixed
the other parameters so that $M_h=120$ GeV and $\cos 2\beta =-1$ as will be the
case in the large $\tb$ limit, $\tb \sim 50$, adopted in this analysis. As can
be seen, even for very small masses of these superparticles, $m_{\chi_1^0} \sim
50$ GeV and $m_{\tilde{e}_{L,R}} \sim 100$ GeV [close to the experimental
bounds from negative searches at LEP2], the cross section hardly reaches the
level of $10^{-2}$ fb for a c.m. energy $\sqrt{s}=500$ GeV and even smaller at
higher energies.  This means that even with the integrated luminosities, ${\cal
L}=500$ fb$^{-1}$, expected at these machines, only a handful of events can be
generated in this process. The couplings $g_{h \tilde{e} \tilde{e}}$ will
therefore be very difficult to measure in this mechanism.  \smallskip

\begin{table}
\begin{center}
\renewcommand{\arraystretch}{1.3}
\begin{tabular}{|c|c|c|c|} \hline
$m_{\chi_1^0}$ [GeV] & $m_{\tilde{e}}$ [GeV] & $\sigma(500)$ 
[fb] & $\sigma(800)$ [fb] \\
\hline 
50 & 100 & 0.010 & 0.005 \\
50 & 200 & 0.001 & 0.001 \\
100 & 105 & 0.005 & 0.005 \\
100 & 200 & 0.0006 & 0.001 \\ \hline 
\end{tabular}
\end{center}
\caption{The cross sections, $\sigma(\sqrt{s})$, for the process $\ee \to h 
\chi_1^0 \chi_1^0$ [in fb] for selected values of $m_{\chi_1^0}$ and 
$m_{\tilde{e}}$ at c.m. energies $\sqrt{s}=500$ GeV and 800 GeV.} 
\end{table}

\subsection*{3. Higgs production in association with selectrons and sneutrinos}

The processes $\ee \to \tilde{e} \tilde{e}^* h$ and $\ee \to \tilde{\nu}_e
\tilde{\nu}_e^* h$ are generated by the diagrams of Fig.~2 where the
neutralinos $\chi_{1, \cdots, 4}^0$ and the charginos $\chi_{1,2}^\pm$ are
exchanged in the $t$--channel, respectively. In the latter case, only the $Z$
boson is exchanged in the $s$--channel diagrams. If the lightest neutralinos
and chargino are higgsino--like, the contributions from the $t$--channel
diagrams are very small since $\chi_{1,2}^0$ and $\chi_1^\pm$ have couplings
proportional to $m_e$ [only the heavier ino states would contribute but the
cross sections are then suppressed since these particles are heavier]. In this
case, the production cross sections are approximately the same as for
associated Higgs production with $\tilde{\mu}$ and $\tilde{\tau}$ [in the case
of no--mixing] and the corresponding sneutrinos, since for these particles
there is no $t$--channel exchange diagram.  If the $\chi_{1,2}^0$ and
$\chi_1^\pm$ particles are gaugino--like all diagrams would contribute [and one
can neglect the contribution of the heavier chargino and neutralino states
which are higgsino--like]. In the mixed gaugino--higgsino region, where the
Higgs boson couplings to neutralinos and charginos are sizeable, the last
diagrams of Fig.~2a has to be taken into account; since we are interested only
in the Higgs--slepton coupling, we will not discuss this region here. 
\vspace*{.7cm}

\begin{picture}(1000,180)(10,0)
\Text(5,180)[]{\bf a)}
\Text(25,130)[]{$e^+$}
\Text(25,170)[]{$e^-$}
\ArrowLine(20,120)(80,120)
\ArrowLine(20,180)(80,180)
\ArrowLine(80,120)(80,180)
\Text(70,150)[]{$\tilde{\chi}$}
\DashLine(80,120)(140,120){4}
\DashLine(80,180)(140,180){4}
\Text(135,130)[]{$\tilde{l}$}
\Text(135,170)[]{$\tilde{l}$}
\DashLine(110,180)(140,150){4}
\Text(130,150)[]{$h$}
\Text(175,130)[]{$e^+$}
\Text(175,170)[]{$e^-$}
\ArrowLine(170,120)(230,120)
\ArrowLine(170,180)(230,180)
\ArrowLine(230,120)(230,180)
\Text(220,150)[]{$\tilde{\chi}$}
\DashLine(230,120)(290,120){4}
\DashLine(230,180)(290,180){4}
\Text(285,130)[]{$\tilde{l}$}
\Text(285,170)[]{$\tilde{l}$}
\DashLine(260,120)(290,150){4}
\Text(280,150)[]{$h$}
\Text(325,130)[]{$e^+$}
\Text(325,170)[]{$e^-$}
\ArrowLine(320,120)(380,120)
\ArrowLine(320,180)(380,180)
\ArrowLine(380,120)(380,180)
\Text(370,150)[]{$\chi$}
\DashLine(380,120)(440,120){4}
\DashLine(380,180)(440,180){4}
\Text(440,130)[]{$\tilde{l}$}
\Text(440,170)[]{$\tilde{l}$}
\DashLine(380,150)(440,150){4}
\Text(425,140)[]{$h$}
\Text(5,80)[]{\bf b)}
\Text(15,30)[]{$e^+$}
\Text(15,70)[]{$e^-$}
\ArrowLine(20,20)(55,50)
\ArrowLine(20,80)(55,50)
\Photon(55,50)(90,50){4}{6}
\Text(70,63)[]{$(\gamma)Z$}
\DashLine(90,50)(135,80){4}
\DashLine(90,50)(135,20){4}
\Text(140,30)[]{$\tilde{l}$}
\Text(140,70)[]{$\tilde{l}$}
\DashLine(110,40)(130,50){4}
\Text(140,50)[]{$h$}
\Text(165,30)[]{$e^+$}
\Text(165,70)[]{$e^-$}
\ArrowLine(170,20)(205,50)
\ArrowLine(170,80)(205,50)
\Photon(205,50)(240,50){4}{6}
\Text(220,63)[]{$(\gamma)Z$}
\DashLine(240,50)(280,20){4}
\DashLine(240,50)(280,80){4}
\Text(280,45)[]{$h$}
\DashLine(265,70)(290,50){4}
\Text(285,30)[]{$\tilde{l}$}
\Text(285,70)[]{$\tilde{l}$}
\Text(315,30)[]{$e^+$}
\Text(315,70)[]{$e^-$}
\ArrowLine(320,20)(355,50)
\ArrowLine(320,80)(355,50)
\Photon(355,50)(390,50){4}{6}
\Text(370,63)[]{$Z$}
\DashLine(390,50)(430,20){4}
\Text(435,30)[]{$h$}
\Photon(390,50)(410,70){4}{6}
\DashLine(410,70)(440,90){4}
\DashLine(410,70)(440,60){4}
\Text(445,90)[]{$\tilde{l}$}
\Text(445,60)[]{$\tilde{l}$}
\end{picture}

\nn Figure 2: The Feynman diagrams contributing to the production of the 
lightest Higgs boson in association with sleptons, $\ee \to \tilde{l}
\tilde{l}^* h$. 
\bigskip

Some values of the cross sections for the associated production processes $\ee
\to \tilde{e} \tilde{e}^* h$, $\ee \to \tilde{\nu}_e \tilde{\nu}_e^* h$ as well
as $\ee \to \tilde{\mu} \tilde{\mu}^* h$ and $\ee \to \tilde{\nu}_\mu
\tilde{\nu}_\mu^*h$ are displayed in Table 2 for selected values of the LSP and
slepton masses at c.m. energies of 500 and 800 GeV. We have summed the cross
sections over possible chiral combinations of final state sleptons and as
previously, have chosen a common mass $m_{\tilde{l}}$ for all sleptons. The
numbers in the cases of the $\ee \to \tilde{e} \tilde{e}^*h$ and $\ee \to
\tilde{\nu}_e \tilde{\nu}_e^* h$ cross sections are for bino--like LSPs; as
mentioned previously; for higgsino--like LSPs, the cross sections are
approximately the same as those for the processes $\ee \to \tilde{\mu}
\tilde{\mu}^*h$ and $\ee \to \tilde{\nu}_\mu \tilde{\nu}_\mu^* h$, 
respectively. 

As can been seen from Table 2, the cross sections for $\tilde{e}, \tilde{\mu}$
and $\tilde{\nu}_\mu$ final states are also very small, below 0.03 fb at the
considered c.m. energies, even for relatively small values of the LSP and
slepton masses, $m_{\chi_1^0}=50$ GeV and $m_{\tilde{l}}=100$ GeV. Only for
$\tilde{\nu}_e \tilde{\nu}_e^* h$ final states, with $m_{\chi_1^0}$ and
$m_{\tilde{\nu}_e}$ values close to the bounds indicated by experimental data,
that the cross sections can reach the level of $\sim 0.2$ fb. This is mainly
due to the large contribution of the chargino mediated $t$--channel diagram
[the charged $e$--$\tilde{\nu}_e$--$\chi_1^\pm$ coupling is stronger than the
neutral $e$--$\tilde{e}$--$\chi_{1,2}^0$ couplings involved in selectron
production] and to the fact that $g^2_{h \tilde{\nu} \tilde{\nu}} \sim 4
g^2_{h\tilde{e} \tilde{e}}$. However, for such small masses, the sneutrino
$\tilde{\nu}_e$ will dominantly decay into invisible final states,
$\tilde{\nu}_e \to \nu_e \chi_1^0$ and thus remains experimentally undetectable
[recall that here, $m_{\chi_1^+} \sim 2 m_{\chi_1^0}$, and the charged visible
decay $\tilde{\nu}_e \to e^\mp \chi_1^\pm$ would be phase--space suppressed].  

Thus, the prospects of measuring the Higgs-slepton couplings in these processes
are rather gloomy, even for the high--luminosities, ${\cal L}=500$ fb$^{-1}$,
expected at the future $\ee$ machines.  

\begin{table}[htbp]
\begin{center}
\renewcommand{\arraystretch}{1.1}
\begin{tabular}{|c|c|c|c|c|} \hline
$\tilde{l}$ & $m_{\chi_1^0}$ [GeV] &$m_{\tilde{l}}$ [GeV] 
& $\sigma(500)$ [fb] & $\sigma(800)$ [fb] \\ \hline
$\tilde{e}$ & 50 & 100 & 0.021 & 0.027 \\
 & 100 & 150 & 0.003 & 0.011 \\
$\tilde{\nu}_e$ & 50 & 100 & 0.127 & 0.195 \\
 & 100 & 150 & 0.006 & 0.055 \\
$\tilde{\mu}$ & 50 & 100 & 0.004 & 0.004 \\
 & 100 & 150 & 0.0004 & 0.002 \\
$\tilde{\nu}_\mu$ & 50 & 100 & 0.0004 & 0.001 \\
 & 100 & 150 & 0.00005 & 0.0006 \\ \hline
\end{tabular}
\end{center}
\caption{The cross sections, $\sigma(\sqrt{s})$, for the processes $\ee \to h 
\tilde{l} \tilde{l}^*$ [in fb] for selected values of the LSP and slepton 
masses at c.m. energies $\sqrt{s}=500$ GeV and 800 GeV.} 
\vspace*{-.5cm}
\end{table}

\subsection*{4. Higgs production in association with staus}

The main reason for the smallness of the cross sections for the processes
discussed in the previous sections is the smallness of the Higgs--slepton
coupling itself. Indeed, compared to the $h$ boson coupling to top squarks,
$g_{h\tilde{t} \tilde{t}} \sim m_t^2/M_Z^2$ in the no--mixing case [i.e. 
$A_t-\mu {\rm cot}\beta \sim 0$], the Higgs coupling to selectrons is one order
of magnitude smaller than its coupling to top squarks, leading to a two--order
of magnitude smaller production cross section $\sigma(\ee \to h \tilde{e}
\tilde{e}^*)$ as compared to $\sigma(\ee \to h\tilde{t}_1 \tilde{t}_1^*)$. 
Since the latter hardly reaches the femtobarn level \cite{hstop}, the results 
for the production rates in the previous sections were to be expected. 

The case of the $\tau$ sleptons is drastically different from the one of the
other sleptons. Indeed, because of the relatively large value of $m_\tau$, the
leading component in the $g_{h \tilde{\tau}_i \tilde{\tau}_i}$ coupling,
eq.~(2), is the one proportional to $\sin 2 \theta_\tau m_\tau (A_\tau - \mu
\tb)$. For large values\footnote{Note that large values of $\mu \sim {\cal O}$
(1 TeV) can be obtained naturally in mSUGRA from the requirement of radiative
electroweak symmetry breaking \cite{DM}, while very large values of $\tb \sim 
{\cal O} (50)$ are favored if one requires Yukawa coupling unification at the 
GUT scale; see Ref.~\cite{mSUGRA}.} of $\mu$ and $\tan \beta$ [or/and extremely
large values of $A_\tau$], the mixing in the $\tilde{\tau}$ sector becomes very
strong, $|\sin 2\theta_\tau| \simeq 1$, leading at the same time, to two 
important consequences: 

$i)$ The mass splitting between the two $\tilde{\tau}$ eigenstates becomes
large, leading to a $\tilde{\tau}_1$ state much lighter than the other
sleptons; the process $\ee \to \tilde{\tau}_1 \tilde{\tau}_1^* h$ will therefore
be more phase space favored than the slepton processes discussed in the
previous section. 

$ii)$ The $h \tilde{\tau}_1 \tilde{\tau}_1$ coupling can be strongly enhanced. 
For instance, for the values $\mu=500$ GeV and $\tb=50$, leading to $|\sin 2
\theta_\tau| \simeq 1$, one has $g_{h \tilde{\tau}_1 \tilde{\tau}_1} \sim 2.5$
compared to $g_{h \tilde{\tau}_1 \tilde{\tau}_1} \sim 0.25$ in the case of
no--mixing. The cross section for $\ee \to \tilde{\tau}_1 \tilde{\tau}_1^* h$ 
can thus be much larger than those involving selectron, smuon and sneutrino
final states. 

The cross section\footnote{Here we will only deal with the continuum cross
section. Another possibility to generate $\tilde{\tau}_1 \tilde{\tau}_1^* h$
final states would be to produce mixed $\tilde{\tau}_1 \tilde{\tau}_2^*$ pairs
in $\ee$ collisions, with the heavier $\tilde{\tau}_2$ decaying into a
$\tilde{\tau}_1h$ final state. This two--step process, however, needs large
enough phase space so that the heavier $\tilde{\tau}_2$ eigenstate can be
produced; in addition the branching ratio for the decay $\tilde{\tau}_2 \to
\tilde{\tau}_1 h$ is small since in the large mixing case, the coupling $g_{h
\tilde{\tau}_1 \tilde{\tau}_2}$, eq.~(2), is close to zero.} for the process
$\ee \to \tilde{\tau}_1 \tilde{\tau}_1^*h$ is similar to that of the associated
production of the $h$ boson with top squarks \cite{hstop} after appropriate
replacements of the couplings, charge and color factors. At high energies and
when the $g_{h \tilde{\tau}_1 \tilde{\tau}_1}$ coupling is large, the cross
section can be approximated by the sole contribution from the photon exchange
diagrams with $h$ emitted from the slepton lines. This is due to the fact that
both of the couplings $g_{Z\tilde{\tau}_1\tilde{\tau}_1}$ and $g_{h\tilde{\tau
}_1\tilde{\tau}_2}$ are proportional, in the large mixing case $|\sin
2\theta_\tau| \to 1$, to $s_W^2 - 1/4$ which is close to zero for $s_W^2 \sim
0.23$, and one can safely neglect the contributions of the $Z$--exchange
diagrams and those involving $\tilde{\tau}_2$ virtual states. In this
case, the Dalitz plot density is given by the very simple formula:
\beq
\frac{ {\rm d} \sigma } { {\rm d} x_1 x_2}
= \frac{\alpha \ \sigma_0 } {16 \pi s_W^2 c_W^2} \, \frac{M_Z^2}{s} \, 
g^2_{h \tilde{\tau}_1\tilde{\tau}_1} 
\left[ \frac{ 1-2x_1  +4 \mu_{\tilde{\tau}} }{(1-x_1)^2} + 
\frac{x_1+x_2 -1+2 \mu_h- 4\mu_{\tilde{\tau}} }{(1-x_1)(1- x_2)} + x_1 
\leftrightarrow x_2 \right] 
\eeq
where $x_{1,2}$ are the reduced energies of the $\tilde{\tau}_1,
\tilde{\tau}_1^*$ final states, $x_{1,2}=2E_{\tilde{\tau}_1, \tilde{\tau}_1^*}
/\sqrt{s}$ and $\mu_i$ the reduced mass squared, $\mu_i = M_i^2/s$;   
$\sigma_0= 4 \pi \alpha^2/3s$ is the QED point--like cross section. 

The  $\ee \to h \tilde{\tau}_1 \tilde{\tau}_1^*$ cross sections in Fig.~3 are
shown\footnote{For the numerical analysis, we have used the complete formula 
for the cross section, including the small contributions from the diagrams 
where 
the $h$ boson is emitted from the $Z$--line and with the exchange of 
$\tilde{\tau}_2$.} as a function of the $\tilde{\tau}_1$ mass for two c.m.
energies, $\sqrt{s}=500$ and 800 GeV. We have fixed the $h$ boson mass to
$M_h=120$ GeV and the SUSY parameters to: $\tb=50, \mu=-A_\tau = 500$ GeV and 
varied the soft--SUSY breaking $\tilde{\tau}$ masses, $m_{\tilde{\tau}_L} 
\simeq m_{\tilde{\tau}_R}$, to vary $m_{\tilde{\tau}_1}$.  
As can be seen, for relatively small $m_{\tilde{\tau}_1}$ values, the cross 
section can exceed $\sim 0.2$ fb, leading to more than 100 events 
for an expected integrated luminosity ${\cal L}=500$ fb$^{-1}$. The cross
section scales quadratically with the parameters $\mu$ and $\tb$ and can
therefore be larger (smaller) when the values of these parameters are increased
(decreased); for instance, for $\mu \sim 1$ TeV and $\tb \sim 50$, $\sigma( 
\ee \to h \tilde{\tau}_1 \tilde{\tau}_1^*)$ reaches the femtobarn level. 

In the case where the main decay mode of the stau would be $\tilde{\tau}_1 \to
\tau \chi_1^0$, the final state would consist of a $b\bar{b}$ pair [since the
main decay mode of the $h$ boson is $h \to b\bar{b}$] peaking at $M_h$ [which
would be precisely measured in the main Higgs production processes], two
tau leptons and a fair amount of missing energy [when the $m_{\tilde{\tau}_1} -
m_{\chi_1^0}$ difference is substantial]:
\beq
\ee \to \tilde{\tau}_1 \tilde{\tau}_1 h \to \tau^+ \tau^- \, + \, b\bar{b} \, 
+ \, \not \hspace*{-1.5mm} E
\eeq
This signal would be not too difficult to detect in the clean environment of
$\ee$ colliders.  A detailed analysis taking into account background and
detection efficiencies, which is beyond the scope of this paper, is
nevertheless required to assess in which part of the MSSM parameter space this
final state can be isolated experimentally.  

\setcounter{figure}{2}
\begin{figure}[htbp]
\vspace*{-3mm}
\begin{center}
\epsfig{file=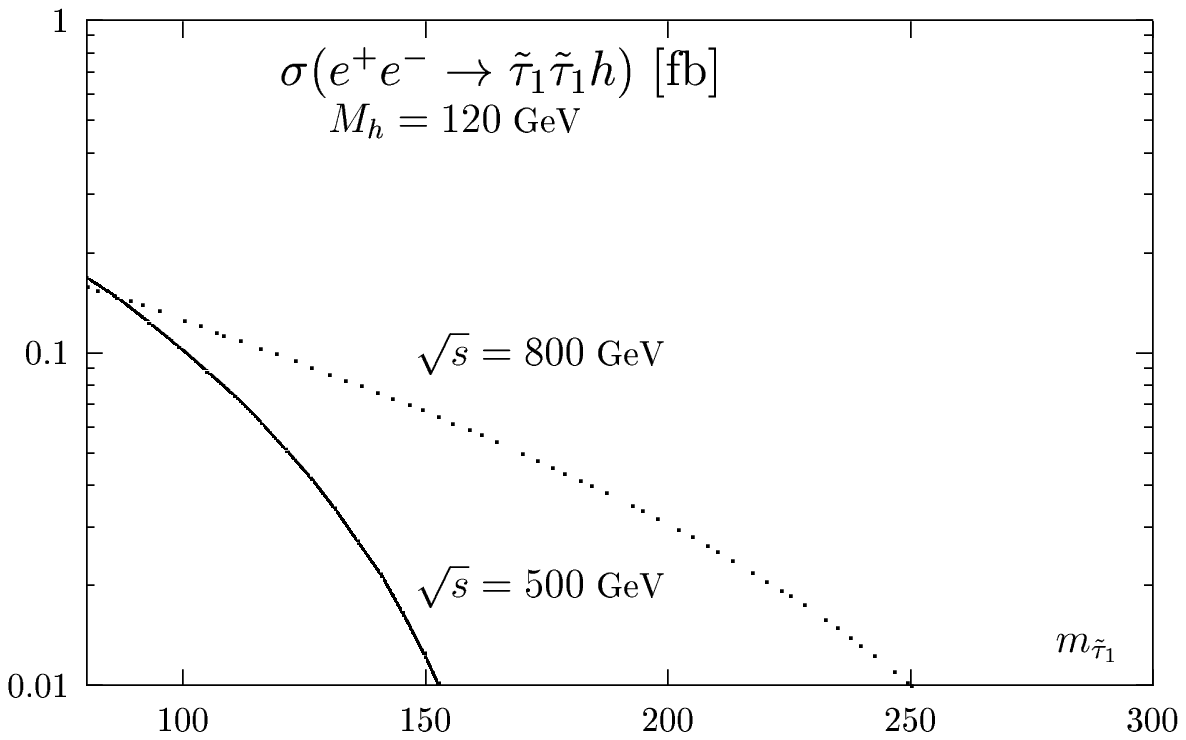,bbllx=140,bblly=450,bburx=480,bbury=700,height=8.0cm}
\end{center}
\vspace*{-1.9cm}
\caption[]{The cross sections for associated $\tilde{\tau}_1 \tilde{\tau}_1^* 
h$ production as a function of $m_{\tilde{\tau}_1}$ in $\ee$ collisions 
at $\sqrt{s}=500$ and 800 GeV; $\mu=-A_\tau=500$ GeV and $\tb=50$.}
\vspace*{-.8cm}
\end{figure}
\begin{figure}[htbp]
\begin{center}
\epsfig{file=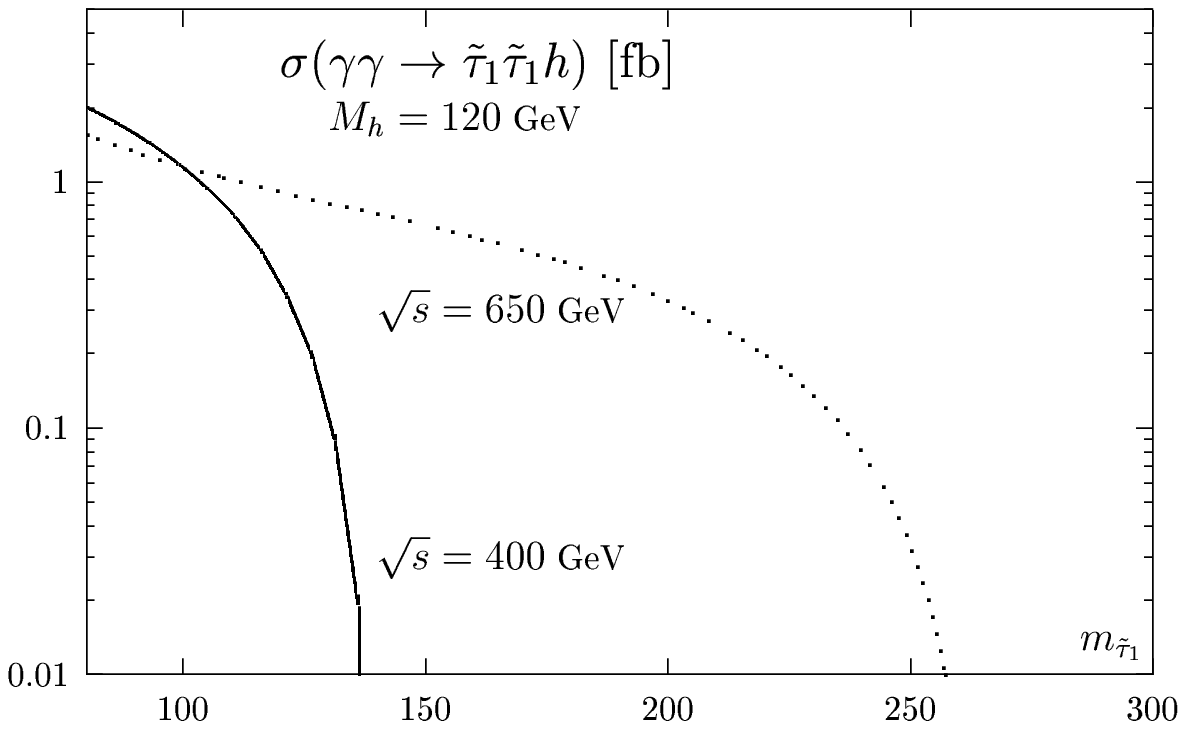,bbllx=140,bblly=450,bburx=480,bbury=700,height=8.0cm}
\end{center}
\vspace*{-1.9cm}
\caption[]{The cross sections for associated $\tilde{\tau}_1 \tilde{\tau}_1^* 
h$ production as a function of $m_{\tilde{\tau}_1}$ in $\gamma \gamma$ 
collisions at $\sqrt{s}_{\gamma \gamma}=400$ and 650 GeV; $\mu=-A_\tau=500$ 
GeV and $\tb=50$.}
\vspace*{-1.5cm}
\end{figure}
\newpage

For completeness, we have also studied the associated production of the $h$
boson with $\tilde{\tau}$ sleptons in photon--photon collisions, $\gamma \gamma
\to h \tilde{\tau}_1 \tilde{\tau}_1^*$, since future $\ee$ linear colliders can
be turned into high--energy $\gamma \gamma$ colliders [with the photons coming
from Compton back--scattering of laser beams] which may have $\sim 80\%$ of the
c.m.~energy and $\sim 50\%$ of the luminosity available at the original $\ee$ 
machine \cite{gamma}.  The process $\gamma \gamma \to h \tilde{\tau}_1 
\tilde{\tau}_1^*$ is generated by two diagrams: one with $\tilde{\tau}_1$ 
exchanged in the $t$--channel and one involving the quartic $\gamma \gamma 
\tilde{\tau}_1 \tilde{\tau}_1^*$ coupling, the $h$ boson being emitted from 
the external and internal slepton lines. 

The cross sections for the subprocess\footnote{The cross section has, in
principle, to be convoluted with the photon spectra; but here for illustration,
we will simply tune the energy and the luminosity of the $\gamma \gamma$
collider at the maximum. Note that, in the expression of the differential
cross section for $\gamma \gamma \to h \tilde{t}_1 \tilde{t}_1^*$ given in 
Ref.~\cite{hstop}, a color factor is missing so that the cross section is three
times larger than shown in the corresponding figure.} are shown in Fig.~4 for
two c.m.  energies, $\sqrt{s}_{\gamma \gamma}=400$ and 650 GeV, with the same
inputs as in Fig.~3.  These are almost an order of magnitude larger than the
corresponding ones at the $\ee$ mode of the collider for relatively small
$m_{\tilde{\tau}_1}$, and reaches the femtobarn level for
$m_{\tilde{\tau}_1}\sim 100$ GeV for both c.m. energies.  For comparable
luminosities of the $\gamma \gamma$ and $\ee$ colliders, and despite the
smaller $\gamma \gamma$ c.m. energy, a sizeable number of events might be
collected for small $m_{\tilde{\tau}_1}$ and large $\tb, \mu$ values.  For
instance, at $\sqrt{s}_{\gamma \gamma} =650$ GeV, stau masses up to
$m_{\tilde{\tau}_1} \sim 250$ GeV can be probed for the inputs of Fig.~4, if
one requires $\sim 50$ events for detectability at a luminosity ${\cal L}_{\ee}
\sim {\cal L}_{\gamma \gamma} \sim 500$ fb$^{-1}$.

\subsection*{5. Conclusions} 

We have studied the associated production of the lightest $h$ boson along with
a neutralino or a slepton pair at future $\ee$ colliders in the context of the
MSSM. The cross sections for these processes are proportional to the
Higgs--slepton couplings and, hence, would allow for their measurements if they 
are large enough.  

It turns out, mainly because of the fact that the couplings of Higgs bosons to
first and second generation charged sleptons as well as to sneutrinos are
rather tiny, the cross sections for these associated production processes are
too small to generate a sufficient number of events, even with the large
luminosity expected at these colliders. An exception might be the process $\ee
\to \tilde{\nu}_e \tilde{\nu}_e^* h$ which, because of the large contribution
from the $t$--channel chargino exchange, might have cross sections at the level
of $\sim 0.2$ fb, for $\chi_1^\pm$ and $\tilde{\nu}_e$ mass values not much
beyond the present experimental bounds.  

In contrast, the couplings of the $h$ boson to $\tilde{\tau}$ pairs can be
sizeable for large values of the parameters $\tan \beta$ and $\mu$, leading to
reasonably large $\ee \to h \tilde{\tau}_1 \tilde{\tau}_1^*$ production cross
sections for not too heavy $\tilde{\tau}_1$, in particular at the $ \gamma
\gamma$ option of the $\ee$ collider. The determination of the $h\tilde{\tau}
\tilde{\tau}$ coupling would provide a very important information on the SUSY
Lagrangian. Indeed, the fact that this coupling is proportional to $\tan
\beta$, can be exploited to measure this fundamental parameter which is,
otherwise [in contrast to the case of the parameter $\mu$, for instance], very
difficult to be determined in other processes when it is rather
large\footnote{Indeed, the parameter $\tb$ can be directly measured only in the
associated production of the pseudoscalar $A$ boson with $b\bar{b}$ pairs
\cite{bbH} for small $A$ masses, $M_{A} \lsim 100$ GeV. The value of $\tb$ is
difficult to determine from chargino/neutralino production \cite{chi} for $\tb
\gsim 10$ since, in this case, the observables depend only on $\cos 2\beta$
which becomes flat for $\beta \to \pi/2$. In sfermion, in particular
$\tilde{\tau}$, production \cite{sfer}, besides the fact that other parameters
[such as the soft--SUSY breaking scalar masses] enter the analysis, one needs
enough phase--space to produce both  $\tilde{\tau}_1$ and $\tilde{\tau}_2$
which have a large mass splitting in the large $\tan \beta$ and $\mu$
scenario.}. 

\newpage

\nn {\bf Acknowledgments:} 
We thank Manuel Drees for discussions. Asesh Datta is 
supported by a MNERT fellowship; A.D. and J.L.K are supported by the 
GDR--Supersym\'etrie and by the European Union under contract 
HPRN-CT-2000-00149.


\begin{thebibliography}{99}

\bibitem{MSSM} For reviews on Supersymmetry and the MSSM see: H. P. Nilles, 
Phys. Rep. 117 (1985) 1; P. Nath, R. Arnowitt and A. Chamseddine, {\it Applied 
N=1 Supergravity}, ICTP series in Theoretical Physics, World Scientific,
Singapore, 1984;  Haber and G. Kane, Phys. Rep. 117 (1985) 75.

\bibitem{HHG} For a review of the Higgs sector in the MSSM, see J.F. Gunion, 
H.E. Haber, G.L. Kane and S. Dawson, {\it The Higgs Hunter's Guide}, 
Addison--Wesley, Reading 1990.

\bibitem{hmass} See M. Carena et al, Nucl. Phys. B580 (2000) 29, and references
therein. 

\bibitem{mSUGRA} For a review, see e.g. M. Drees and S.P. Martin, 
hep-ph/9504324.

\bibitem{gunion} J.F. Gunion and H.E. Haber, Nucl. Phys. B272 (1986) 1; (E) 
hep-ph/9301205. 

\bibitem{DM} See e.g,  T. Falk, Phys. Lett. B456 (1999) 171.

\bibitem{hstop0} A. Djouadi, J.L. Kneur and G. Moultaka, Phys. Rev. Lett. 80 
(1998) 1830; G. B\'elanger, F. Boudjema, T. Kon 
and V. Lafage, Eur. Phys. J. C9 (1999) 511; A. Dedes and S. Moretti, Eur.
Phys. J. C10 (1999) 515. 

\bibitem{hstop} A. Djouadi, J.L. Kneur and G. Moultaka, Nucl. Phys. B569 
(2000) 53. 

\bibitem{lep2} For a recent compilation of LEP2 results, see T. Junk (for the 
LEP Collaborations), hep-ex/0101015.  

\bibitem{eeH} For reviews of Higgs production at future $\ee$ colliders, see: 
P.M. Zerwas (ed.), Proceedings Workshop {\it Physics with $\ee$ Linear 
Colliders}, hep-ph/9605437; A. Djouadi, Int. J. Mod. Phys. A10 (1995) 1; 
E. Accomando, Phys. Rept. 299 (1998) 1. 

\bibitem{gamma} For a recent review, see V. Telnov, hep-ex/0010033.  

\bibitem{bbH} A.~Djouadi, J.~Kalinowski and P.M.~Zerwas, Mod. Phys. Lett.
A7 (1992) 1765 and Z. Phys. C54 (1992) 255; J. Kalinowski and M. Krawczyk, 
Phys. Lett. B361 (1995) 66; M. Berggren, R. Keranen and A. Sopczak, 
hep-ph/9911344. 


\bibitem{chi} J. Feng and M. Strassler, Phys. Rev. D55 (1997) 1326;
J.L. Kneur and G. Moultaka, Phys. Rev. D59 (1999) 015005; 
S.Y. Choi et al, Eur. Phys. J. C7 (1999) 123 and Eur. Phys. J. C14 (2000) 535;
G. Moortgat-Pick et al., Eur. Phys. J. C18 (2000) 379. 

\bibitem{sfer} 
M. Nojiri, K. Fujii and T. Tsukamoto, Phys. Rev. D54 (1996) 6756;  
A. Bartl et al., hep-ph/0002115 and  hep-ph/0010018.  

\end{thebibliography}
\end{document}